%% file: main.tex
\documentclass[conference]{IEEEtran}
\usepackage{cite}
\usepackage{amsmath,amssymb,amsfonts}
\usepackage{algorithmic}
\usepackage{graphicx}
\usepackage{textcomp}
\usepackage{xcolor}
\usepackage[hyphens]{url}

\usepackage{amsmath}
\usepackage{graphicx}
\usepackage{caption}
\usepackage{subcaption}
\usepackage{xcolor}
\usepackage{color, colortbl}
\usepackage{listings}
\usepackage{mathtools}
\usepackage{hyperref} 

\DeclarePairedDelimiter\ceil{\lceil}{\rceil}

\definecolor{codegreen}{rgb}{0,0.6,0}
\definecolor{codecustomblue}{rgb}{0.05,0.11,0.98}
\definecolor{codegray}{rgb}{0.5,0.5,0.5}
\definecolor{codepurple}{rgb}{0.58,0,0.82}
\definecolor{backcolour}{rgb}{0.97,0.97,0.94}

\definecolor{key-color}{rgb}{0.41, 0.25, 0.03}
\definecolor{key-colors2}{rgb}{0.61, 0.45, 0.13}

\lstdefinestyle{mystyle}
{
    float=tp,
    language = C++,
    basicstyle = {\small,\ttfamily },
    backgroundcolor=\color{backcolour},
    numbersep=0pt,
    commentstyle=\color{codegreen},
    stringstyle = {\color{string-color}},
    keywordstyle = {\color{blue}},
    keywordstyle = [2]{\color{codecustomblue}},
    keywordstyle = [3]{\color{key-color}},
    keywordstyle = [4]{\color{codepurple}},
    keywordstyle = [5]{\color{key-colors2}},
    otherkeywords = {;,<<,>>,++},
    morekeywords = [2]{fms_dt},
    morekeywords = [3]{FCM_Skeleton},
    morekeywords = [4]{INT8_DTYPE,BUFFER_SIZE},
    morekeywords = [5]{communication_buffer},
    numbers=left,
}

\def\BibTeX{{\rm B\kern-.05em{\sc i\kern-.025em b}\kern-.08em
    T\kern-.1667em\lower.7ex\hbox{E}\kern-.125emX}}

\title{Fusing Depthwise and Pointwise Convolutions for Efficient Inference on GPUs}

\begin{document}

\author{
\IEEEauthorblockN{Fareed Qararyah, Muhammad Waqar Azhar, Mohammad Ali Maleki, Pedro Trancoso}
\IEEEauthorblockA{\textit{Department of Computer Science and Engineering} \\
\textit{Chalmers University of Technology, Gothenburg, Sweden}\\
\{qarayah, waqarm, mohammad.ali.maleki, ppedro\}@chalmers.se}
}

\maketitle
\thispagestyle{plain}
\pagestyle{plain}


\begin{abstract}

\input{sections/0_abstract}
\end{abstract}

\begin{IEEEkeywords}
depthwise convolution, pointwise convolution, CNN, vision transformer, layer fusion, GPU
\end{IEEEkeywords}

\input{sections/1_Introduction}
\input{sections/2_background}
\input{sections/4_1_fcms}
\input{sections/4_2_fuse_estimator}
\input{sections/5_experimental_setup}
\input{sections/6_evaluation}
\input{sections/7_related_work}
\input{sections/8_conclusion}



\bibliographystyle{IEEEtranS}
\bibliography{main}

\end{document}

%% file: sections/0_abstract.tex
Depthwise and pointwise convolutions have fewer parameters and perform fewer operations than standard convolutions. As a result, they have become increasingly used in various compact DNNs, including convolutional neural networks (CNNs) and vision transformers (ViTs). However, they have a lower compute-to-memory-access ratio than standard convolutions, making their memory accesses often the performance bottleneck.\\
This paper explores fusing depthwise and pointwise convolutions to overcome the memory access bottleneck. The focus is on fusing these operators on GPUs. The prior art on GPU-based fusion suffers from one or more of the following: (1) fusing either a convolution with an element-wise or multiple non-convolutional operators, (2) not explicitly optimizing for memory accesses, (3) not supporting depthwise convolutions. This paper proposes Fused Convolutional Modules (FCMs), a set of novel fused depthwise and pointwise GPU kernels. FCMs significantly reduce pointwise and depthwise convolutions memory accesses, improving execution time and energy efficiency. To evaluate the trade-offs associated with fusion and determine which convolutions are beneficial to fuse and the optimal FCM parameters, we propose FusePlanner. FusePlanner consists of cost models to estimate the memory accesses of depthwise, pointwise, and FCM kernels given GPU characteristics. Our experiments on three GPUs using representative CNNs and ViTs demonstrate that FCMs save up to 83\% of the memory accesses and achieve speedups of up to 3.7x compared to cuDNN. Complete model implementations of various CNNs using our modules outperform TVMs' achieving speedups of up to 1.8x and saving up to two-thirds of the energy. FCM and FusePlanner implementations are open source: \href{https://github.com/fqararyah/Fusing_DW_and_PW_on_GPUs}{https://github.com/fqararyah/Fusing\_DW\_and\_PW\_on\_GPUs}

%% file: sections/1_introduction.tex
\section{Introduction}

Convolutions are core operators in many Deep Learning (DL) models, including Convolutional Neural Networks (CNNs)~\cite{krizhevsky2017imagenet,howard2017mobilenets,he2016deep}, Vision Transformers (ViTs)~\cite{wu2021cvt, xiao2021early, guo2022cmt, yu2021glance,dai2021coatnet,yuan2021incorporating}, and Graph Convolutional Networks (GCNs)~\cite{zhang2019graph,schlichtkrull2018modeling}. Splitting a standard convolution into depthwise (DW) and pointwise (PW) convolutions reduces the model size and operation count~\cite{chollet2017xception, howard2017mobilenets}. To give an example, XCeption CNN, which uses DW and PW convolutions, has an accuracy that surpasses ResNet-152's~\cite{he2016deep} despite being roughly three times smaller~\cite{chollet2017xception}. Hence, DW and PW convolutions are increasingly replacing standard convolutions in designing compact models that achieve state-of-the-art accuracy~\cite{chollet2017xception,howard2017mobilenets, tan2019efficientnet, so2021primer, wu2021cvt, yuan2021incorporating}. 

 \begin{figure}[!t]
         \centering
         \captionsetup{justification=centering}
         \includegraphics[width=\linewidth]{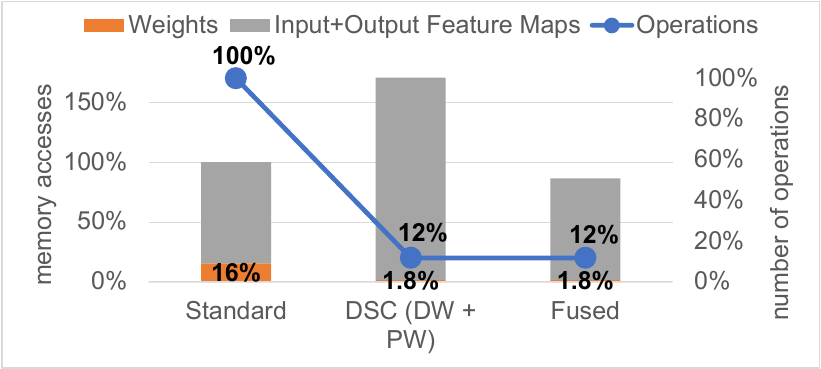}
         \caption{Operation count and memory accesses of a standard, DSC (DW+PW), and fused convolutions. The example is taken from MobileNet. All values are normalized to the standard convolution values}
         \label{fig:intro_fig}
\end{figure}

Figure~\ref{fig:intro_fig} demonstrates the effect of splitting a standard convolution into depthwise separable convolution (DSC)~\cite{howard2017mobilenets}, \textit{i.e.} DW plus PW, on the operation count, weights, and input and output sizes. The DW and PW convolutions require fewer weights and perform fewer operations than standard convolutions. However, their combined inputs and outputs are larger. The net result is having fewer operations but more memory accesses. In other words, DW and PW are more often memory-bound compared to standard convolutions~\cite{lu2021optimizing}. As a result, their memory accesses form a performance bottleneck on most of the commonly used accelerators. 


Operator fusion, or layer fusion, reduces off-chip memory accesses considerably compared to traditional layer-by-layer execution. In layer-by-layer execution, the convolution processes its inputs completely and writes the results to the main memory. However, fusing layers allows the intermediate results to be processed at finer granularity while on-chip~\cite{zhang2022full,alwani2016fused,cai2021optimus,jeong2023pin}. In the example in Figure~\ref{fig:intro_fig}, fusion saves $50\%$ of the off-chip memory accesses of the DW and PW convolutions. Fusing a convolution with an element-wise operation like normalization and non-linearity is a common optimization applied by DL compilers like TVM~\cite{chen2018tvm} and DNNVM~\cite{xing2019dnnvm}. However, due to complex input-output dependencies among convolutions, fusing multiple convolutions could incur numerous redundant computations or memory accesses~\cite{alwani2016fused}. Nonetheless, the prior art has demonstrated that when handling the trade-offs properly, fusing convolutions is beneficial on custom accelerators~\cite{shen2017maximizing, gao2019tangram, qararyah2024efficient, alwani2016fused, xiao2017exploring, umuroglu2017finn, blott2018finn, zheng2020efficient, jeong2023pin, cai2021optimus, wei2018tgpa, xing2019dnnvm, li2016high, yang2023isosceles,zhang2022full, olyaiy2021accelerating}.

As GPUs have been key accelerators in the resurgence of DL~\cite{krizhevsky2017imagenet} and are the most widely-supported accelerators by various DL frameworks~\cite{bergstra2011theano, jia2014caffe, abadi2016tensorflow, paszke2019pytorch}, they are an ideal target of various optimizations including layer fusion. However, when it comes to fusing multiple convolutions, GPU programming abstractions and memory level access constraints make managing cross-convolution dependencies challenging~\cite{alwani2016fused}. 


The prior art on layer fusion suffers from one or more of the following limitations. First, on GPUs, they either consider fusing a convolution and an element-wise or multiple non-convolutional operators~\cite{chen2018tvm, jia2020enabling, li2016optimizing, dong2018characterizing}, do not explicitly model and optimize memory accesses, or do not support depthwise convolutions~\cite{zheng2023chimera, waeijen2021convfusion}. Second, the work targeting other accelerators is either tightly coupled to a specific architecture or assumes complete hardware flexibility like that offered by FPGAs~\cite{li2016high,alwani2016fused,umuroglu2017finn,wei2018tgpa, xiao2017exploring}.


In this paper, we propose \emph{Fused Convolutional Modules (FCMs)}, a set of novel fused GPU kernels of DW and PW convolutions. FCMs reduce global memory access considerably leading to improved latency and energy efficiency. To evaluate fusion trade-offs and decide when fusion gains outweigh its overheads, we propose \emph{FusePlanner}. Given a set of DW and PW convolutions and GPU characteristics, \emph{FusePlanner's} cost models estimate the memory accesses of depthwise, pointwise, and FCM kernels. FusePlanner explores the feasible FCMs and layer-by-layer implementations and suggests one that minimizes memory access. FCMs can be used as library routines, and with FusePlanner they can be used to derive complete CNN execution plans. Using FusePlanner-suggested FCM and layer-by-layer implementations, we implement and evaluate convolutional layers of state-of-the-art CNNs and ViTs on three GPUs. We compare our implementation with CuDNN~\cite{chetlur2014cudnn} based implementations. Moreover, we compare the performance of full implementations of the CNNs, based on FusePlanner-suggested FCM and layer-by-layer kernels, to a DL compiler (TVM)~\cite{chen2018tvm} optimized implementations. Our contributions are as follows:

\begin{itemize}
    \item  We propose Fused Convolutional Modules (FCMs), a set of novel GPU kernels comprising DW and PW convolutions. FCMs mitigate these convolutions' memory access bottleneck leading to low-latency and energy-efficient execution. 
    \item We propose \emph{FusePlanner},  FusePlanner consists of cost models that estimate global memory accesses of DW, PW, and FCM kernels given a GPU architecture. \emph{FusePlanner} decides which layers benefit from fusion and the implementation parameters that minimize global memory accesses.
    \item We evaluate FCMs by comparing their performance to custom, and standard DL library-based (cuDNN) convolution kernels from representative CNNs and ViTs on multiple GPUs. We also compare end-to-end implementations of the CNNs utilizing FCMs and FusePlanner-suggested layer-by-layer implementations to TVM-optimized models.
\end{itemize}

The proposed FCMs achieve up to \emph{$1.8x$} speedup over custom layer-by-layer implementations and up to \emph{$3.7x$} over the best cuDNN implementations using representative CNNs and ViTs. FCMs save up to 83\% of the global memory accesses compared to CuDNN. End-to-end implementations of four CNNs using the proposed kernels achieve up to \emph{$1.8x$} speedup compared to TVM implementations and save up to two-thirds of the energy per inference.

%% file: sections/2_background.tex
\section{Background and Motivation}
\label{sec:background}

\subsection{CNNs and ViTs}
Convolutional neural networks (CNNs) are feed-forward DNNs~\cite{lecun2015deep}. 
As the name suggests, the main layers in a CNN are the convolutional layers. A convolutional layer has a set of trainable parameters or \emph{weights} grouped into \emph{filters}. The filters are applied to multi-dimensional arrays of input or intermediate results, extracting their embedded features~\cite{lecun2010convolutional}. The inputs of a layer are known as input feature maps (\emph{IFMs}) and the outputs as output feature maps (\emph{OFMs}). Feature Maps (\emph{FMs}), or activations, refer to both IFMs and OFMs. FMs comprise a set of 2D slices known as \emph{channels}.

Transformer models are based on a self-attention mechanism that learns the relationships between elements of a sequence~\cite{vaswani2017attention}. In vision transformers (ViTs), \emph{self-attention} allows modeling contextual information of the full image and long-range dependencies both in space and time~\cite{khan2022transformers}. This paper focuses on convolutional ViTs that combine self-attention with convolutions~\cite{wu2021cvt,so2021primer,yuan2021incorporating}.

\subsection{Depthwise and pointwise convolutions}
\label{sec:back_pw_dw}

 \begin{figure}[!htbp]
         \centering
         \captionsetup{justification=centering}
         \includegraphics[width=\linewidth]{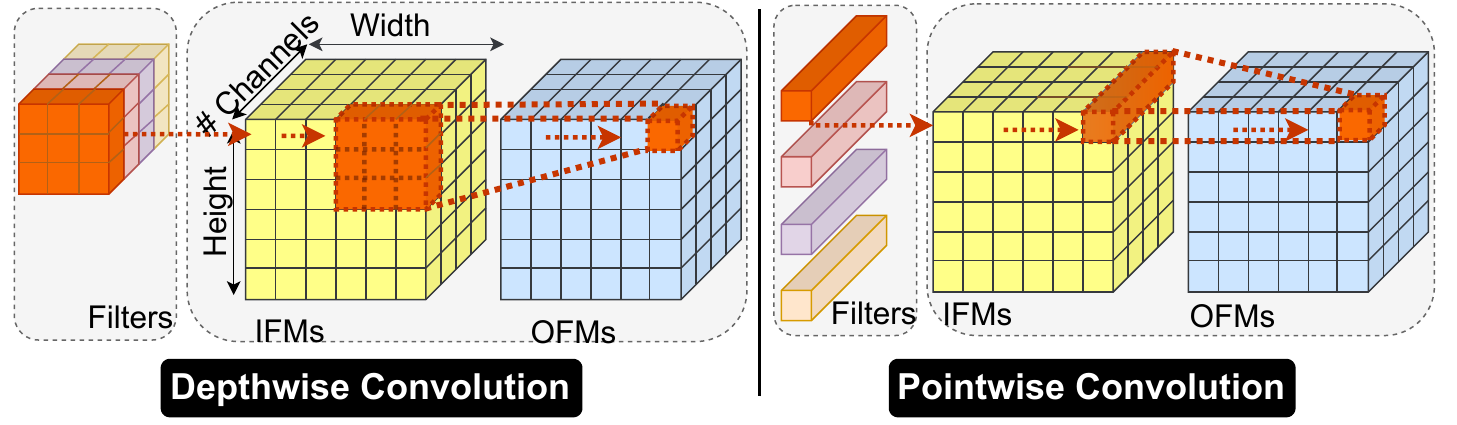}
         \caption{Depthwise and pointwise convolutions}
         \label{fig:dw_pw}
\end{figure}

Depthwise (DW) and Pointwise (PW) convolutions optimize DNNs' size-accuracy trade-off~\cite{chollet2017xception, howard2017mobilenets, sandler2018mobilenetv2, tan2019efficientnet, zhang2018shufflenet, cai2018proxylessnas, so2021primer, wu2021cvt}. They decouple the spatial and cross-channel correlations~\cite{chollet2017xception}. 
As Figure~\ref{fig:dw_pw} shows, DW convolution is applied to the spatial dimensions,\emph{i.e.} width, and height, where one filter is applied to a single channel. PW convolution is applied to the cross-channel dimension, where its $1 \times 1$ filters span over all channels. DW and PW convolutions are combined in various ways to build efficient modules or blocks, including \emph{Depthwise Separable Convolutions (DSC)} and inverted residual with linear bottlenecks, or \emph{inverted residuals} for short,~\cite{chollet2017xception,howard2017mobilenets, sandler2018mobilenetv2, tan2019efficientnet}. The DSC is composed of a DW followed by a PW layer. The inverted residuals comprise three convolutional layers: PW-DW-PW.

\begin{figure*}[hbt!]
 \centering
 \begin{subfigure}[t]{0.39\textwidth}
         \centering
         \captionsetup{justification=centering}
         \includegraphics[width=\textwidth]{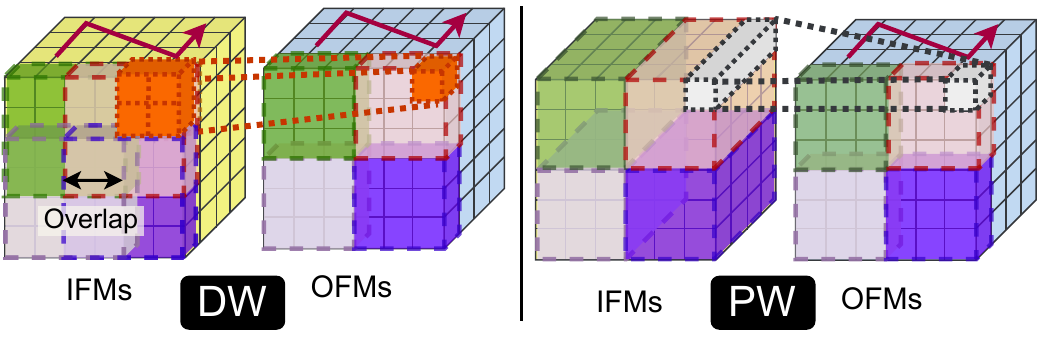}
         \caption{Tiled DW and PW convolution}
         \label{fig:pw_dw_tiled}
     \end{subfigure}
      \hfill
 \begin{subfigure}[t]{0.59\textwidth}
         \centering
         \captionsetup{justification=centering}
         \includegraphics[width=\textwidth]{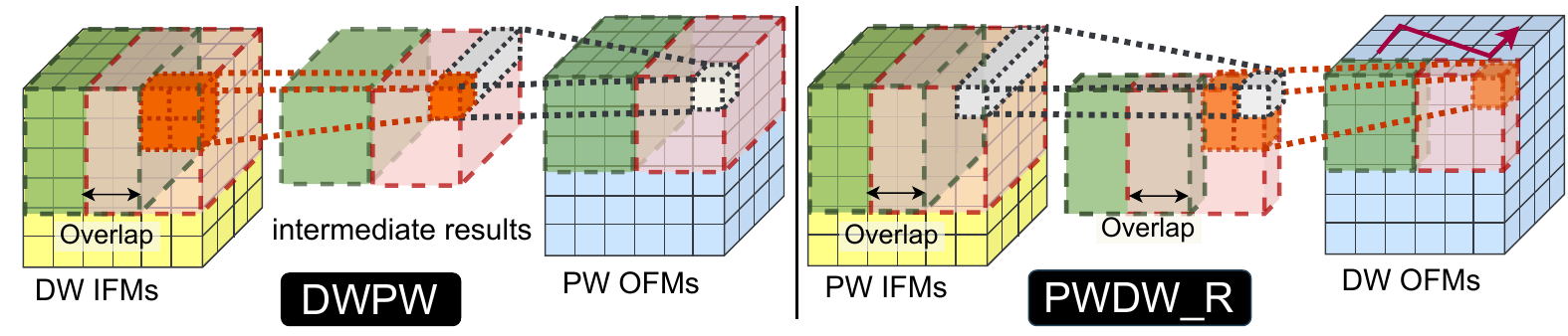}
         \caption{Two fusion cases: DW followed by PW (DWPW), and PW followed by DW (PWDW\_R)}
         \label{fig:dw_pw_fusion}
     \end{subfigure}
\caption{\centering Tiled Layer-by-Layer and fused convolutions}  
\label{fig:tvm_speedup}
\end{figure*} 

\subsection{CUDA-Capable GPU architecture and programming model overview}
\label{subsec:gpu_arch}
A GPU architecture consists of a scalable array of streaming multiprocessors (SMs)~\cite{nickolls2008scalable}. An SM is a Single-Instruction-Multiple-Thread (SIMT) architecture that runs groups of parallel threads called \emph{warps} in a lockstep fashion. A CUDA kernel is processed by a \emph{grid} of threads. The grid consists of a set of \emph{thread blocks}, threads in a block run on the same SM. GPU has a memory hierarchy of multiple levels with different access constraints. Each thread has private local registers. Each SM has a low-latency L1 cache, and a variable-sized portion of that cache is configurable to serve as programmer-managed \emph{shared memory}. The shared memory is visible to all threads in a block and has the same lifetime of the block. The rest of the memory levels are globally accessible by threads of the entire CUDA kernel.

\subsection{Fusing Convolutions on GPUs}

\label{subsec:back_fusing}
There are multiple algorithms to implement convolution on GPU. We focus on the direct convolution implementation and use it as the basis for the layer-by-layer and the fused kernels. This is because other algorithms, including Winograde and FFT, require filters of greater than $1 \times 1$ width and height, so they are not applicable for PW convolution~\cite{lavin2016fast}. Moreover, Winograde, FFT, and GEMM optimize the computation at the cost of more memory bandwidth requirements, which does not suit PW and DW convolutions. 

Figure~\ref{fig:pw_dw_tiled} shows a simplified example of tiled DW and PW convolutions. For simplicity, each weight tile has only one filter and computes a 2d tile of $3 \times 3$ of the OFMs. The DW convolution has $2 \times 2$ filters and $4 \times 4$ tile of the IFMs. On a GPU, assuming an Output Stationary (OS) dataflow, each OFM tile gets assigned to a thread block that runs on one of the GPU's Streaming Multiprocessors (SMs). In the layer-by-layer execution, each layer is implemented as one or more CUDA kernels that process the IFMs and produce the complete OFMs. Because the SMs' L1/shared memory contents do not outlive a single kernel, all the OFMs must be written back and cannot be reused by the next layer. Note that because L1/shared memory is private to an SM, the overlap regions among the IFM tiles, in the case of DW, must be loaded multiple times depending on the number of tiles sharing them.

Figure~\ref{fig:dw_pw_fusion} shows two fusion examples. The first example (DWPW) depicts a DW fused with its following PW, and the second (PWDW\_R) shows a PW fused with a following DW. The \_R indicates that this fusion entails redundant computations, as explained at the end of the section. The fused layers are implemented as a single kernel. On the one hand, unlike the layer-by-layer, the OFMs of the first layer, which are intermediate results when fusing, can be directly reused while in the L1/shared memory. This reduces the global memory access.
On the other hand, the fused implementation has its own constraints and overheads. First, fusing convolutions enlarges the working set compared to the layer-by-layer implementation. In the layer-by-layer case, the working set consists of three tiles: OFMs, IFMs, and filter tiles. In the fused case, there are five tiles: IFMs of the first layer tile, OFMs of the second layer tile, two tiles of both layers' filters, and a tile of the intermediate results exchanged between the two layers. Note that we show one filter in the figures for simplicity, in practice a filter tile may contain hundreds or thousands of filters. As more tiles compete for the L1/shared memory, each has a smaller share. Smaller tiles lead to more overlapping, less reuse, and more frequent access to the global memory. Second, certain fusion cases restrict the viable tile sizes. For example, the PW layer in the DWPW fusion case in Figure~\ref{fig:dw_pw_fusion} requires at least one element of each channel of the intermediate results to produce one valid output. Consequently, the intermediate results and the input tiles must contain all the channels. In other words, a DW tiling similar to the one shown in Figure~\ref{fig:pw_dw_tiled} is not feasible. Third, the values located at the overlap regions of intermediate results tiles must be redundantly computed. PWDW\_R in Figure~\ref{fig:dw_pw_fusion} shows an example of this. Unlike the overlaps in the IFMs, the values at the overlap in the intermediate results do not exist before the fused kernel starts. They must be computed independently by the SMs computing the overlapping tiles.

Searching for fused implementations of PW and DW convolutions that minimize memory accesses, and consequently mitigate or overcome their bottlenecks requires evaluating the gains and overheads of the feasible fusions compared to layer-by-layer execution. We propose Fused implementations of PW and DW convolutions and cost models that evaluate the discussed overheads and suggest implementations that minimize the global memory accesses. 

%% file: sections/4_1_fcms.tex
\section{Fused Convolutional Modules (FCMs)}
     
\label{sec:fcms}

\subsection{FCMs overview}

\begin{figure}[!h]
         \centering
         \captionsetup{justification=centering}
         \includegraphics[width=\linewidth]{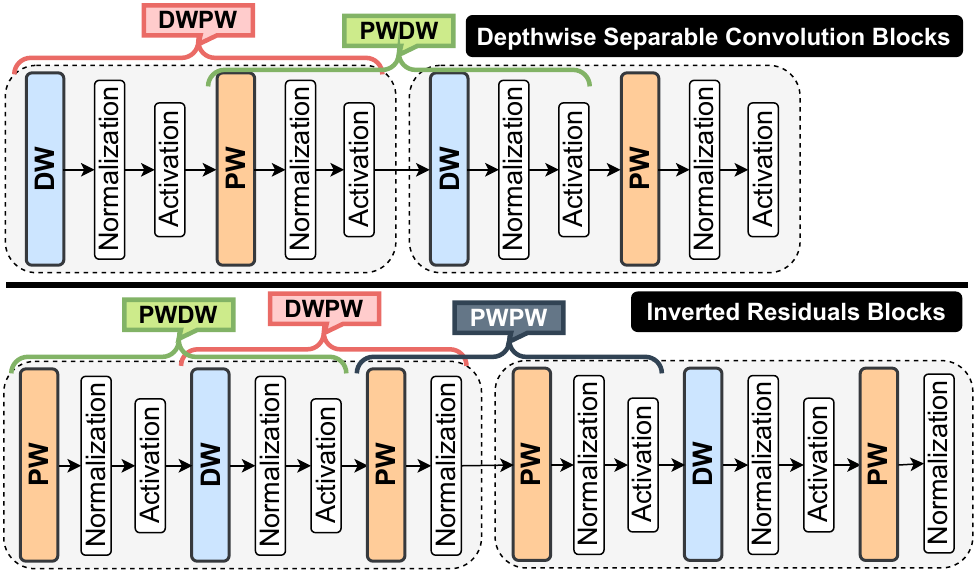}
         \caption{Possible FCMs in DNNs composed of Depthwise Separable Convolutions and Inverted Residuals blocks}
         \label{fig:fcm_combinations}
\end{figure}

DW and PW convolutions are commonly found in DNNs, e.g. CNNs and ViTs, in the form of \emph{depthwise separable convolutions (DSC)}, or \emph{inverted residuals} (Section~\ref{sec:background}). Figure~\ref{fig:fcm_combinations} shows a sequence of two DSC blocks and a sequence of two inverted residuals. It depicts the three possible PW and DW combinations. Fused Convolutional Modules (FCMs) target such combinations, which are \textbf{DWPW}, \textbf{PWDW}, and \textbf{PWPW}. The PWDW FCM has two variants, one that requires redundant computations (\textbf{PWDW\_R} shown in Figure ~\ref{fig:dw_pw_fusion}), and one that does not (\textbf{PWDW}). The PWDW does not require redundant computations if there is no tiling across the width and height of an IFM. An FCM combines up to 6 layers, two convolutional layers, and the normalization and activation layers following each. As Figure~\ref{fig:fcm_combinations} shows, FCMs fuse layers of a single separable convolution or inverted residual blocks, or layers belonging to two consecutive blocks. All the FCM kernels adopt the efficient \emph{Output Stationary - Local Weight Stationary (OS-LWS)} dataflow~\cite{venkatesan2019magnet}.

\subsection{FCM kernel structure}

\begin{lstlisting}[label={lis:psuedo_kernel},firstnumber=1,language=C, style=mystyle, frame=tb,basicstyle=\footnotesize, caption={FCM kernels skeleton}]
__global__ void FCM_Skeleton(/*Parameters*/) {
    /********************part 1********************/
    __shared__ fms_dt commBuffer[BUFFER_SIZE];
    //Other declarations
    /********************part 2********************/
    // Prefetch fused layers weight tiles
    if (/*Thread ID in loader thread IDs*/) {
        //Prefetch weights to shared memory or
        //registers
    }
    // Synchronize
    /********************part 3********************/
    // First layer core
    if (/*Thread ID in Conv1 thread IDs*/) {
        //Compute Conv-Norm-Activation
        //Pack and write to commBuffer
    }
    // Synchronize
    /********************part 4********************/
    // Second layer core
    if (/*Thread ID in Conv2 thread IDs*/) {
        //Load second layer IFM tile from commBuffer
        //Compute Conv-Norm-Activation
        //Pack and write back to the OFMs
    }
}
\end{lstlisting}


Listing~\ref{lis:psuedo_kernel} highlights the main parts of the skeleton of an FCM kernel. The skeleton is divided into four main parts. \emph{Part1} (lines 2-4) contains the declaration of the buffer used to communicate between the first set of layers (convolution-normalization-activation) and the second (\emph{commBuffer}). The buffer is stored in an SM's shared memory. Shared memory banks are organized such that consecutive words map to consecutive banks. The access patterns to these banks are crucial to the kernel performance. To fully utilize these banks' throughput, the data layout is selected based on the FCM layers implementation to always have a linear addressing with a stride of one, a conflict-free access pattern.

\emph{Part2} (lines 6-11) contains the prefetching of layer weights. The weights are fetched ahead of computation in two scenarios. The first scenario is when the implementation of either FCM's two convolutions does not access the weights contiguously by default due to a mismatch between the convolution dataflow and loop ordering, and the data layout of the weights buffer~\cite{li2016optimizing}. In such cases, separating the weights load from the computation allows to load weights contiguously. The second scenario is when warp-level primitives are used. We use a warp-level primitive (\emph{\_\_shfl\_sync}) to exchange the weights between threads through registers rather than shared memory.

In \emph{parts 3 and 4} (lines 13-24), if the weights have been fetched in part 2, they are now loaded from the shared memory or shuffled from other threads registers; otherwise, they are loaded from global memory. Then, a fused convolution-normalization-activation operation is applied. The implementations currently support both INT8 and FP32 data types. In the case of INT8, \emph{\_\_dp4a CUDA intrinsic} four-way dot product with 32-bit accumulate is used. The first convolution-normalization-
activation of the FCM computes a tile of the intermediate results and writes it to the shared \emph{commBuffer}. Then, the fused convolution-normalization-activation part reads the intermediate results from the \emph{commBuffer}, computes, and writes back the FCM output to the OFMs buffer. Synchronization is necessary between these two parts as different threads may participate in each part. When using INT8, every four results are grouped, or packed, into one 32-bit integer before writing to any buffer. The weights are also packed, weight packing is done offline since the weights do not change in inference.


%% file: sections/4_2_fuse_estimator.tex
\section{FusePlanner}
\label{sec:fuseestimator}

\begin{figure}[!htbp]
         \centering
         \captionsetup{justification=centering}
         \includegraphics[width=\linewidth]{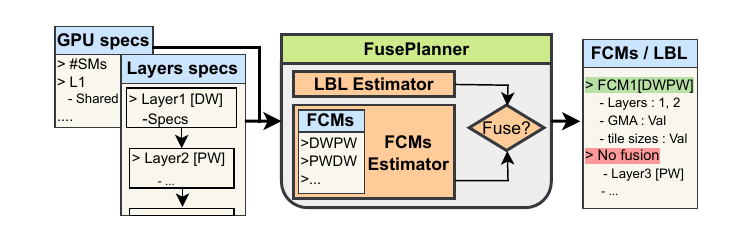}
         \caption{FusePlanner overview}
         \label{fig:fuseestimator}
\end{figure}

FusePlanner aims to identify the FCMs and layer-by-layer implementations that minimize global memory access, given a set of DW and PW layers and GPU specifications. Figure~\ref{fig:fuseestimator} shows an overview of FusePlanner. It takes as inputs: (1) GPU number of SMs, L1 size, and the portion configurable as shared memory; and (2) a DAG representing a model or set of layers, their weight and FM specifications, and the layers connectivity. We currently support generating model DAGs from Tensorflow. FusePlanner has two main components, \emph{layer-by-layer global memory access estimator} and \emph{FCMs global memory access estimator}. FusePlanner does a first pass over the layers and estimates their minimum global memory access using the layer-by-layer global memory access estimator. After that, it examines all the possible fusions and evaluates their global memory access using the FCMs global memory access estimator. Based on the layer-by-layer and FCM estimates, FusePlanner outputs: (1) which layers are to be fused and which are not, (2) which FCMs to use, and (3) the tiling that minimizes the global memory access in each case.

\subsection{Layer-by-layer global memory access estimator}
\label{sec:lbl_estimate}

We propose fast and simple cost models to explore the search space for implementation parameters that minimize global memory access efficiently. To construct a simple cost model, we make two assumptions that prune the search space by excluding implementations that do not perform well on GPUs. First, the data layout guarantees that threads in a warp access consecutive memory locations and that the memory transactions are naturally aligned~\cite{nickolls2008scalable}. Second, the implementation follows an Output Stationary-Local Weight Stationary (OS-LWS) dataflow~\cite{venkatesan2019magnet} and guarantees that the partial sums stay in registers and that only the final results are written to the memory. To guarantee that, all the weights and IFM elements needed to produce one OFM element must be in the same tile (Section~\ref{subsec:back_fusing}). To prove the effectiveness of this approach experimentally, we show that our layer-by-layer implementations outperform CuDNN (Section~\ref{subsec:vs_cudnn}).

The global memory accesses of kernels that meet the two discussed assumptions are estimated using Equations~\ref{eq:pw_gma} and \ref{eq:dw_gma}. Where the \emph{overlap} (described in Section~\ref{subsec:back_fusing}) is obtained using Equation~\ref{eq:tiles_overlapping}, the postfix \emph{GMA} stands for global memory access, \emph{Sz} stands for size, \emph{W} for width, \emph{H} for height, \emph{D} for depth, and \emph{HW} for $height \times width$. As the equations show, the OFMs are written once, because the dataflow (OS-LWS) is a variant of OS. In the PW case, each weight tile is convolved with all IFM tiles, and each IFM tile is convolved with all filters. Hence, weights and IFMs memory accesses depend on each other's tiling. In the DW case, as at least one filter slice must be assigned to each SM (to guarantee assumption 2), there are no weight tiles splitting filters' height and width. As a result, the IFM elements, except the overlapping, are read only once. FusePlanner explores the tile sizes that meet two constraints. The first constraint is that the tiles fit into the L1 cache to avoid misses and redundant loading. Note that the subset of these tiles stored on the shared memory portion of the cache must also fit within that portion. However, the implementation must guarantee that; this is why it is not expressed in the equations. The second is that the number of OFM tiles is greater than or equal to the number of GPU SMs. Having more OFM tiles than the number of GPU SMs ensures that the GPU resources are not underutilized.


\begin{equation}
\label{eq:tiles_overlapping}
\small
\begin{split}
\text{\emph{Overlap}} =\ &(\ceil{ \frac{\text{\emph{ChannelW}}}{\text{\emph{TileW}}} } - 1) \times\ (\text{\emph{FilterW}} - \text{\emph{Strides}}) \times\ \\
& \text{\emph{ChannelH}} \ + (\ceil{\frac{\text{\emph{ChannelH}}}{\text{\emph{TileH}}} } - 1)\\
&\times\ (\text{\emph{FilterH}} - \text{\emph{Strides}}) \times\ \text{\emph{ChannelW}}
\end{split}
\end{equation}


\begin{equation}
\label{eq:pw_gma}
\small
\begin{split}
\text{\emph{PwGMA}} =\ &\ceil{ \frac{\text{\emph{WeightsSz}}}{\text{\emph{WeightsTileSz}}} }\ \times\ \text{\emph{IFMsSz}}\ + \text{\emph{OFMsSz}}\ +\\
&\ceil{ \frac{\text{\emph{OFMsSz}}}{\text{\emph{OFMsTileSz}}} } \times\ \text{\emph{WeightsSz}}\\
\text{\emph{where}}\ & \text{\emph{L1Sz}}\ \geq\ \text{\emph{IFMsTileSz}}\ +\ \text{\emph{OFMsTileSz}}\ +\\
&\text{\emph{WeightsTileSz}}\\
& \text{\emph{and}}\ \#\text{\emph{OFMsTiles}}\geq \#\text{\emph{SMs}}\\
\end{split}
\end{equation}

\begin{equation}
\label{eq:dw_gma}
\small
\begin{split}
\text{\emph{DwGMA}} =\ & 2\ \times\ \text{\emph{IFMsD}}\ \times \ \text{\emph{Overlap}}\ +\ \text{\emph{IFMsSz}}\ +\\
&\text{\emph{OFMsSz}}\ +\ \ceil{\frac{\text{\emph{OFMsHW}}}{\text{\emph{OFMsTileHW}}}}\ \times \text{\emph{WeightsSz}}\\
\text{\emph{where}}\ & \text{\emph{L1Sz}}\ \geq\ \text{\emph{IFMsTileSz}}\ +\ \text{\emph{OFMsTileSz}}\ +\\
&\text{\emph{WeightsTileSz}}\\
& \text{\emph{and}}\ \#\text{\emph{OFMsTiles}}\geq \#\text{\emph{SMs}}\\
\end{split}
\end{equation}




\subsection{FCMs global memory access estimator}
\label{subsec:fused_dm}

Estimating an FCM global memory access is based on Equations~\ref{eq:pw_gma} and \ref{eq:dw_gma}, with two key differences. First, neither the OFMs of the first convolutional layer of an FCM nor the IFMs of its second contribute to the global memory accesses. This is because they are now intermediate results communicated through the communication buffer. Secondly, the accesses of each of the two convolutional layers are affected by the other. Equation~\ref{eq:pwdw_gma} shows an example of FCM's global memory access estimation, a PWDW FCM in this case. The equation again assumes that the fused kernel meets the two assumptions described in the previous section. The equation shows the mentioned two key differences. First, neither the OFMs of the PW nor the IFMs of the DW layer contribute to the global memory accesses. Secondly, the accesses to the first layer's IFMs depend on both layers' weights tiles because the OFMs of the first layer are not written to the global memory. Hence, when the second layer needs them, they must be recomputed, which requires redundant loading of the corresponding IFM elements. Finally, the overlap accesses depend on both layers' IFMs. As the equation shows, the overall overlap is obtained by multiplying the PW IFMs depth, rather than the DW IFMs depth, by DW IFM overlap. This is due to the same reason, \textit{i.e.} OFMs of the FCM's first layer are not written to the global memory, and the overlap in the second layer's IFM elements are obtained by loading the first layer's IFMs and recomputing. The equations of the other FCMs are constructed from the PW and DW Equations~\ref{eq:pw_gma}, and~\ref{eq:dw_gma} similarly.

\begin{equation}
\label{eq:pwdw_gma}
\small
\begin{split}
\text{\emph{PwDwGMA}} =\ &(2 \times \text{\emph{PwIFMsD}} \times\ \text{\emph{DwOverlap}}\ +\ \text{\emph{PwIFMsSz}})\\
&\times\ \text{\emph{max}}(\\
&\ceil{ \frac{\text{\emph{PwWeightsSz}}}{\text{\emph{PwWeightsTileSz}}}}\ ,\ceil{\frac{\text{\emph{DwWeightsSz}}}{\text{\emph{DwWeightsTileSz}}}})\ + \\
&\ceil{ \frac{\text{\emph{DwOFMsSz}}}{\text{\emph{DwOFMsTileSz}}} }\ \times\ \text{\emph{PwWeightsSz}}\ +\\
& \ceil{ \frac{\text{\emph{DwOFMsHW}}}{\text{\emph{DwOFMsTileHW}}} }\ \times\ \text{\emph{DwWeightsSz}}\ \\
\text{\emph{where}}\ \text{\emph{L1Sz}}\ \geq & \ \text{\emph{PwIFMsTileSz}}\ +\ \text{\emph{DwOFMsTileSz}}\ +\\
&\text{\emph{PwWeightsTileSz}} +\\ &\text{\emph{DwWeightsTileSz}}\ +\ \text{\emph{CommBufferSz}} \\
& and\ \#\text{\emph{FCM}}\_\text{\emph{OFMsTiles}}\geq \#\text{\emph{SMs}}\\
\end{split}
\end{equation}



The first constraint in Equation~\ref{eq:pwdw_gma} is more restrictive in FCMs than layer-by-layer, as five tiles rather than three compete for the L1. And as the equation shows, simply having smaller tiles is not always a solution as it may increase the overall memory accesses. The effect of having two weight tiles per SM, compared to one tile in the layer-by-layer case, is not crucial when fusing DW and PW since DW weights are much smaller than PWs' in most cases. However, the effect becomes considerable when two PW layers are fused. That is why PWPW fusion is less likely when the weights use FP32 compared to INT8 (Table~\ref{tab:layers_characteristics}).

FusePlanner explores all tile sizes that meet the constraints in Equations~\ref{eq:pw_gma},~\ref{eq:dw_gma}, and \ref{eq:pwdw_gma} and identifies the ones that minimize the global memory accesses for the layer-by-layer and all the possible FCM cases. The explored tile sizes are restricted to multiples of the warp size to avoid resource underutilization. FusePlanner suggests fusing, when there is an FCM for which the minimum estimated global memory accesses are less than those of its constituting layers. Otherwise, a layer-by-layer implementation is suggested.





%% file: sections/5_experimental_setup.tex
\section{Experimental Setup}
\label{sec:setup}

\subsection{Evaluation system}
\label{sec:system}
We use three GPUs, listed in Table \ref{tab:GPUs}, with different resources representing different points in the compute continuum. We refer to them as \emph{GTX}, \emph{RTX}, and \emph{Orin} in the rest of the paper. CUDA-11.6 is used. CUDA events API is used to measure the execution time, and \emph{nvidia-smi} utility to measure the power consumption on GTX and RTX and \emph{tegrastats} on Orin. \emph{NVIDIA Nsight Compute} is used to quantify accesses to all memory levels and their throughput and to categorize kernels into compute and memory-bound.

\begin{table}
\centering
\caption{Used GPUs specifications}
\setlength\tabcolsep{2pt}
\resizebox{0.9\columnwidth}{!}{
\begin{tabular}{|l|c|c|c|c|c|c|} \hline
\rowcolor{gray!10}
GPU                               &Compute    &\#SM &CUDA&L1/shared &L2   &Off-chip \\ 
\rowcolor{gray!10}                &Capability &     &cores&(KB)     &(MB) &Memory   \\ \hline
\cellcolor{gray!10} \textbf{GTX}-1660      &7.5        &22   &1408 &96       &1.5  &GDDR5    \\ \hline
\cellcolor{gray!10}\textbf{RTX}-A4000      &8.6        &128  &6144 &128      &4    &GDDR6    \\ \hline
\cellcolor{gray!10}Jetson AGX \textbf{Orin}&8.7        &16   &2048 &192      &4    &LPDDR5   \\ \hline
\end{tabular}
}
\label{tab:GPUs}
\end{table}

\begin{table*}
\centering
\caption{Fusion cases and their ratios of redundant computations. F1-F12 using FP32, and F1\_8-F12\_8 using INT8.}
\setlength\tabcolsep{2pt}
\resizebox{1.6\columnwidth}{!}{
\begin{tabular}{|l|c|c|c|c|c|c|c|c|c|c|c|c|}
\hline
\rowcolor{gray!10}
DNN&Mob\_v1&Mob\_v1 &Mob\_v2 &Mob\_v2 & XCe & XCe &Prox &Prox &CeiT &CeiT &CMT &CMT \\
\hline
\cellcolor{gray!10}FP32&F1&F2&F3&F4&F5&F6&F7&F8&F9&F10&F11&F12 \\
\cline{2-13}
\cellcolor{gray!10}&
PWDW\_R&
PWDW\_R&
DWPW&
PWDW\_R&
PWDW\_R&
PWDW\_R&
DWPW&
PWDW\_R&
PWDW&
PWDW\_R&
PWDW&
PWDW\_R \\
\cline{2-13}
\cellcolor{gray!10}&
7\%&
13\%&
-&
18\%&
4\%&
7\%&
-&
10\%&
-&
16\%&
-&
13\%
 \\
\hline
\cellcolor{gray!10}INT8&F1\_8&F2\_8&F3\_8&F4\_8&F5\_8&F6\_8&F7\_8&F8\_8&F9\_8&F10\_8&F11\_8&F12\_8 \\
\cline{2-13}
\cellcolor{gray!10}&
DWPW&
PWDW&
DWPW&
PWPW&
DWPW&
PWDW\_R&
DWPW&
PWPW&
PWDW&
PWDW&
PWPW&
PWDW \\
\cline{2-13}
\cellcolor{gray!10}&
-&
-&
-&
-&
-&
15\%&
-&
-&
-&
-&
-&
-
 \\
\hline
\end{tabular}
}
\label{tab:layers_characteristics}
\end{table*}

\subsection{Workloads}

We evaluate the proposed modules using PW and DW convolutions from four representative CNN models and two ViTs. These are MobileNet (\textbf{Mob\_v1})~\cite{howard2017mobilenets}, MobileNetV2 (\textbf{Mob\_v2})~\cite{sandler2018mobilenetv2}, XCeption (\textbf{XCe})~\cite{chollet2017xception}, ProxylessNAS (\textbf{Prox})~\cite{cai2018proxylessnas}, \textbf{CeiT}~\cite{yuan2021incorporating}, and \textbf{CMT}~\cite{guo2022cmt}. The evaluation is done with FP32 and INT8, the original and the commonly used precision in inference, respectively. We do a fine-grained evaluation using pairs of layers, or fusion cases, from these DNNs that FusePlanner suggested. Table~\ref{tab:layers_characteristics} lists these fusion cases, from which DNNs they are selected, and the ratios of redundant computations if there are any. These cases represent the scenarios where FusePlanner suggests the same fusion type across the three GPUs. A fusion case may occur in a DNN once or multiple times. This is because DNNs usually contain replicated blocks composed of layers of the same hyper-parameters. For example, F1\_8 in the INT8 case represents the second and third layers of Mob\_v1, but F2\_8 represents five pairs of layers (pairs located between layers 14-24). The fused layers, consequently the FCMs, in the case of INT8 are not necessarily the same in the case of FP32. For example, F1\_8 is different from F1 in FP32. Changing the bit-width changes the tile sizes causing FusePlanner to make different choices.

\subsection{Baselines}
\label{sce:baselines}

To demonstrate the effect of fusion on global memory access reduction and performance improvement of PW and DW convolutions, we implement and compare the FCMs and layer-by-layer kernels (\emph{LBL}). To isolate the fusion effect, the LBL kernels have similar dataflow and access patterns to their fused counterpart. We compare the FCMs and LBL kernels to \emph{cuDNN}~\cite{chetlur2014cudnn}. cuDNN gives the flexibility of choosing the convolution algorithms, we compare against three cuDNN algorithms that yielded the best performance on our workloads, namely GEMM, IMPLICIT\_GEMM, and IMPLICIT\_PRECOMP\_GEMM. We also do end-to-end evaluations where the four CNNs, Mob\_v1, Mob\_v2, XCe, and Prox, are fully implemented using our FCMs and LBL kernels and compared with \emph{TVM}~\cite{chen2018tvm}. TVM is an open-source and widely used end-to-end deep learning compiler. We use cuDNN as the backend of our TVM implementation to maintain consistency. TVM applies layer fusion between convolution and non-convolution layers as a core optimization. Not all TVM models ran successfully on all the GPUs after applying TVM's offline graph optimizations. Hence we ran auto-tuning for 20 iterations with the hardware in the loop which was sufficient for all models. TVM offers several tuning heuristic options, in our experiments, we compare our results with the best TVM heuristic in each experiment. Note that cuDNN and TVM implementations that we compare against fuse a single convolutional layer with the normalization and activation layers following it. However, we refer to their execution as layer-by-layer (LBL) since they do not fuse multiple convolutional layers.

%% file: sections/6_evaluation.tex
\section{Evaluation}
\label{sec:evaluation}

\subsection{Fusion effect: comparing FCMs to LBL}

\begin{table}

\caption{\centering Categorizing the FP32 LBL and FCMs into compute and memory-bound based on Roofline model analysis. C means compute-bound and M means memory-bound.}
\resizebox{\columnwidth}{!}{
\setlength\tabcolsep{2pt}
\begin{tabular}{|l|c|c|c|c|c|c|c|c|c|c|c|c|c|}
\hline
\rowcolor{gray!10}&&F1&F2&F3&F4&F5&F6&F7&F8&F9&F10&F11&F12 \\
\hline
\cellcolor{gray!10}GTX&\cellcolor{gray!10}LBL&
\cellcolor{green!20}M, M&
\cellcolor{red!30}C, M&
\cellcolor{green!20}M, M&
\cellcolor{green!20}M, M&
\cellcolor{red!30}C, M&
\cellcolor{red!30}C, M&
\cellcolor{green!20}M, M&
\cellcolor{green!20}M, M&
\cellcolor{red!30}C, M&
\cellcolor{red!30}C, M&
\cellcolor{red!30}C, M&
\cellcolor{red!30}C, M\\
\cline{2-14}
\cellcolor{gray!10}&\cellcolor{gray!10}FCM&
C&
C&
M&
C&
C&
C&
M&
C&
C&
C&
C&
C\\
\hline
\cellcolor{gray!10}RTX&\cellcolor{gray!10}LBL&
\cellcolor{green!20}M, M&
\cellcolor{red!30}C, M&
\cellcolor{green!20}M, M&
\cellcolor{green!20}M, M&
\cellcolor{red!30}C, M&
\cellcolor{red!30}C, M&
\cellcolor{green!20}M, M&
\cellcolor{green!20}M, M&
\cellcolor{red!30}C, M&
\cellcolor{red!30}C, M&
\cellcolor{green!20}M, M&
\cellcolor{red!30}C, M\\
\cline{2-14}
\cellcolor{gray!10}&\cellcolor{gray!10}FCM&
M&
C&
M&
M&
C&
C&
M&
M&
C&
C&
C&
C\\
\hline
\end{tabular}
}
\label{tab:compute_mem_bound}
\end{table}

\begin{figure}[!htbp]
         \centering
         \captionsetup{justification=centering}
         \includegraphics[width=\linewidth]{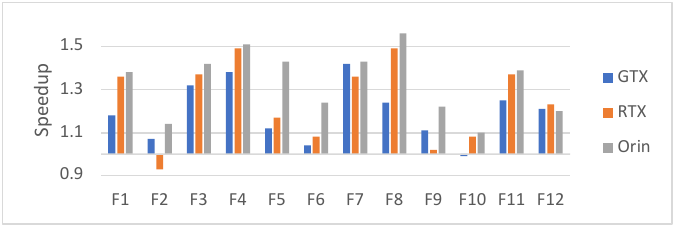}
         \caption{Speedup of FCMs over LBL using FP32}
         \label{fig:speedup_fn_fp32}
\end{figure}
\begin{figure}[!htbp]
         \centering
         \captionsetup{justification=centering}
         \includegraphics[width=\linewidth]{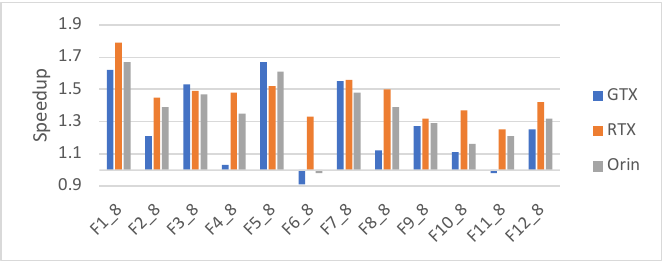}
         \caption{Speedup of FCMs over LBL using INT8}
         \label{fig:speedup_fn_int8}
\end{figure}

This section analyzes the effect of fusion using various workloads, two precisions, and three GPUs. Figures ~\ref{fig:speedup_fn_fp32} and~\ref{fig:speedup_fn_int8} show the speedup achieved as a result of the fusion in the 24 cases (Table~\ref{tab:layers_characteristics}) on the three GPUs using FP32 and INT8 precision. FCMs outperform LBL in 67 out of the 72 experiments. The maximum achieved speedup using FP32 is $1.6x$ in the case of F8 on Orin and $1.8x$ using INT8 in the case of F1\_8 on RTX. The average speedups are $1.3x$ and $1.4x$ using FP32 and INT8 respectively. Orin and RTX have the best average speedups using FP32 and INT8, respectively. GTX has the lowest speedups in both cases. In the rest of this section, we discuss three factors that determine the speedup achieved by FCMs. We then explore the fusion effect on different GPUs and finally with different precisions.

\begin{figure*}[!htbp]
         \centering
         \captionsetup{justification=centering}
         \includegraphics[width=\textwidth]{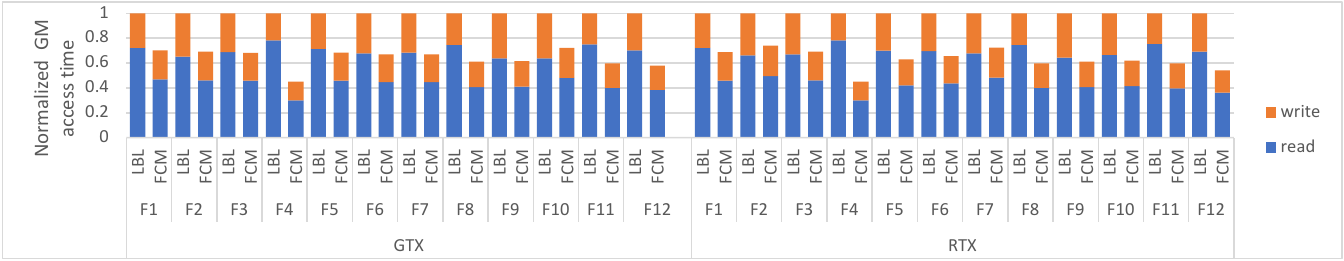}
    \caption{ Global memory access time of FCMs compared to LBL using FP32. The breakdown shows loads and stores' contribution. All values are normalized to the total global memory access time of the LBL case}
    \label{fig:mat_fn_fp32}
\end{figure*}

\textbf{Factors that determine FCMs speedup:} The first factor governing the effect of fusion on speedup is whether the fused kernels are \textbf{memory- or compute-bound}. In general, the memory access reduction is translated to speedup for memory-bound but not for compute-bound kernels. Table~\ref{tab:compute_mem_bound} shows which kernels fall under each category. In the case of RTX, F1, 3, 4, 7, 8, and 11, which consist of two-memory-bound layers, have higher speedups than the rest (Figure~\ref{fig:speedup_fn_fp32}). The average speedup of these six layers is $1.4x$ compared to $1.1x$ for the other six. The same applies to GTX, the five cases where both layers are memory-bound have an average speedup of $1.3x$ compared to $1.1x$ for the rest. Speedups among layers within each category, compute-bound and memory-bound, are determined by the \textbf{amount of reduction in memory access time}. Figure~\ref{fig:mat_fn_fp32} shows the global memory access time of both FCM and LBL executions normalized to that of LBL. For example, among the RTX six FCMs where both layers are memory-bound, F4 has the highest memory access reduction resulting in the highest speedup. And F12 has the highest memory access reduction resulting in the highest speedup among the six cases where at least one layer is compute-bound. However, there are some exceptions. The third factor, the existence of \textbf{redundant computations}, explains these exceptions. For example, on GTX, F7 has the highest speedup among FCMs where both layers are memory-bound even though F4 experiences a larger reduction of the memory access time. This is because, unlike F7, F4 has 18\% redundant computations (Table~\ref{tab:layers_characteristics}).

There are two cases where there is a non-negligible slowdown. These cases are F2 on RTX and F6\_8 on GTX. The three factors discussed explain this slowdown. For example, in the case of F2 on RTX (Figure~\ref{fig:speedup_fn_fp32}) not both layers are memory-bound (Table~\ref{tab:compute_mem_bound}), the memory access reduction is relatively low (Figure~\ref{fig:mat_fn_fp32}), and there are redundant computations (Table~\ref{tab:layers_characteristics}). 

\textbf{Speedup across GPUs:} Orin and RTX have higher speedups than GTX. Moreover, out of the five cases where FCMs do not have speedup over LBL, three are on GTX compared to one on Orin and One on RTX. One reason is that GTX has the smallest L1/shared memory per SM (Table~\ref{tab:GPUs}). This gives less room to the tiles competing on this memory, including the communication buffer (Section~\ref{sec:fuseestimator}). Another reason is that GTX has fewer CUDA cores (Table~\ref{tab:GPUs}), making fewer LBL kernels memory-bound. Table~\ref{tab:compute_mem_bound} shows examples of that. First, RTX has 6 out of 12 cases where both kernels are memory-bound compared to 5 out of 12 on GTX. Secondly, among the 5 memory-bound cases on both GPUs, 3 remain memory-bound after fusion on RTX, namely F1, F4, and F8. As long as a kernel is memory-bound, reducing global memory access time is, ideally, purely translated into speedup. However, on GTX, these turn into compute-bound when fused, meaning that their performance benefited from the global memory access reduction only partially. To summarise, our method identifies fusions that are advantageous across different GPUs. However, the fusion effect on performance varies depending on the GPU compute and memory resources.

\textbf{Speedup and precision:} Note that the FP32 FCMs are different from the INT8 ones, but both precision's FCMs are representative of layers selected by FusePlanner given the same DNN models (Section~\ref{sec:setup}). Hence, we here comment on the general trends rather than having case-by-case comparisons. The maximum and the average speedups are higher using INT8 compared to FP32. This is mainly because reducing the data size allows the L1/shared memory to fit larger tiles (Section~\ref{sec:fuseestimator}). This in turn permits fusion types that are not feasible in FP32. For example, as Table~\ref{tab:layers_characteristics} shows, the dominant FCM using FP32 is PWDW\_R which requires redundant computation, but in INT8 there is only one PWDW\_R. In other words, most INT8 fusions do not have redundant computations making the fusion effect more apparent.

\subsection{Comparison with CuDNN}
\label{subsec:vs_cudnn}

 \begin{figure*}[!htbp]
         \centering
         \captionsetup{justification=centering}
         \includegraphics[width=\textwidth]{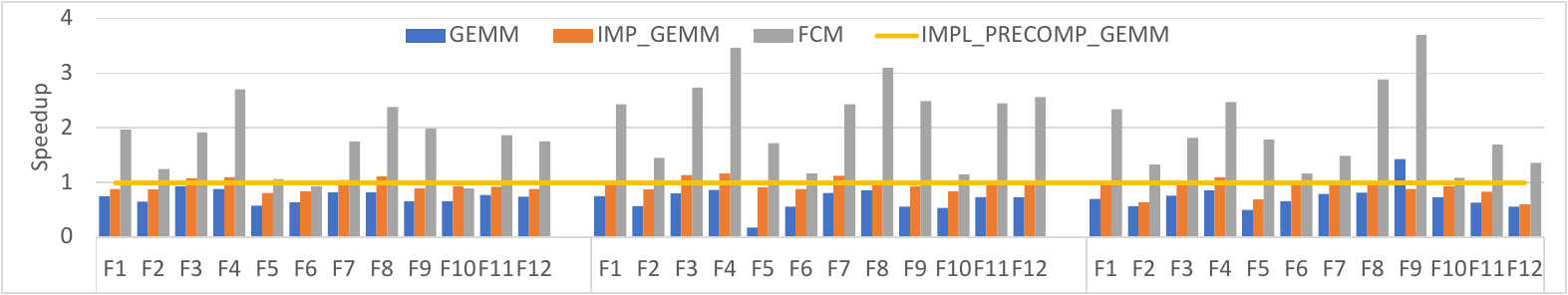}
         \caption{Speedup of FCMs over CuDNN using FP32. All values are normalized to IMPL\_PRECOMP\_GEMM.}
         \label{fig:speedup_cudnn_fp32}
\end{figure*}

Figure~\ref{fig:speedup_cudnn_fp32} shows a comparison between FCMs and CuDNN, and the speedup of FCMs over the best cuDNN algorithms, namely IMPL\_PRECOMP\_GEM. The maximum speedup is $3.7x$, and the average is $2x$. Our LBL implementations also outperform CuDNN in all cases and achieve a maximum speedup of $3x$, and an average speedup of $1.5x$.  When comparing cuDNN implementations, the implicit GEMM implementations outperform direct GEMM. Implicit GEMMs do not explicitly form the matrix that holds the input data resulting in fewer memory accesses. Compared to implicit IMPL\_PRECOMP\_GEM, the best among the three CuDNN algorithms, our LBL implementations save up to 63\% of the global memory accesses, and FCMs save up to 83\%. Generally speaking, the results trend is similar to the one discussed in the previous section. For example, in the cases where the pair is composed of memory-bound layers on RTX and GTX, namely F1, 3, 4, 7, and 8, FCMs have relatively high speedups over cuDNN. In addition, both RTX and Orin experience higher speedups compared to GTX. Both FCMs and LBL outperform cuDNN in the INT8 case as well. This is implicitly shown in the next section as we compare our results with TVM implementations configured to use cuDNN in the backend.

\begin{figure*}[!hbpt]
 \centering
       \begin{subfigure}[t]{\columnwidth}
         \centering
         \captionsetup{justification=centering}
         \includegraphics[width=\textwidth]{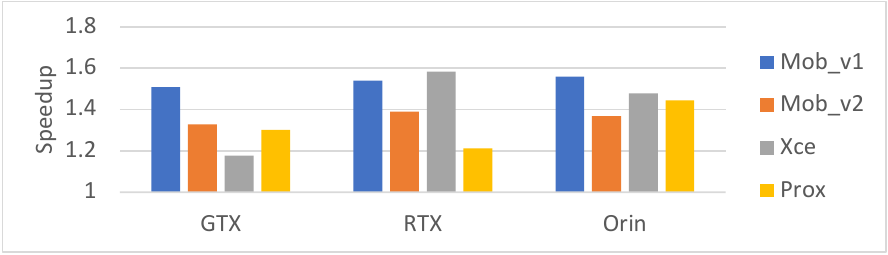}
         \caption{Speedup over TVM using FP32.}
         \label{fig:tvm_speedup_fp32}
     \end{subfigure}
     \hfill
      \begin{subfigure}[t]{\columnwidth}
         \centering
         \captionsetup{justification=centering}
         \includegraphics[width=\textwidth]{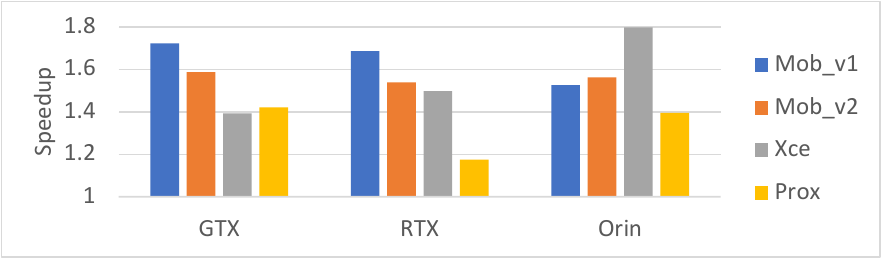}
         \caption{Speedup over TVM using INT8}
         \label{fig:tvm_speedup_int8}
     \end{subfigure}
\caption{\centering Speedup of CNN implementations using FCMs and FusePlanner suggested LBL kernels over TVMs'}  
\label{fig:tvm_speedup}
\end{figure*} 

\begin{figure*}[!hbpt]
 \centering
      \begin{subfigure}[t]{\columnwidth}
         \centering
         \captionsetup{justification=centering}
         \includegraphics[width=\textwidth]{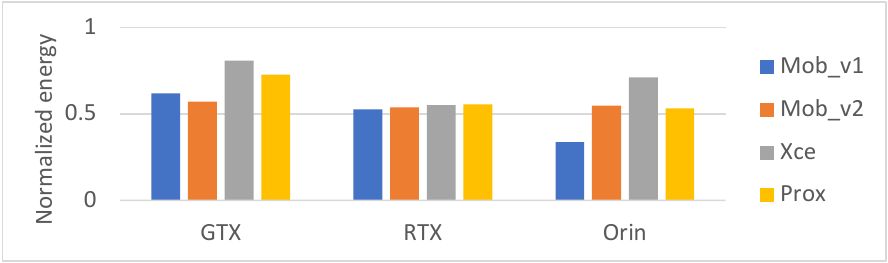}
         \caption{Energy per inference using FP32 normalized to TVM}
         \label{fig:tvm_energy_fp32}
     \end{subfigure}
     \hfill
     \begin{subfigure}[t]{\columnwidth}
         \centering
         \captionsetup{justification=centering}
         \includegraphics[width=\textwidth]{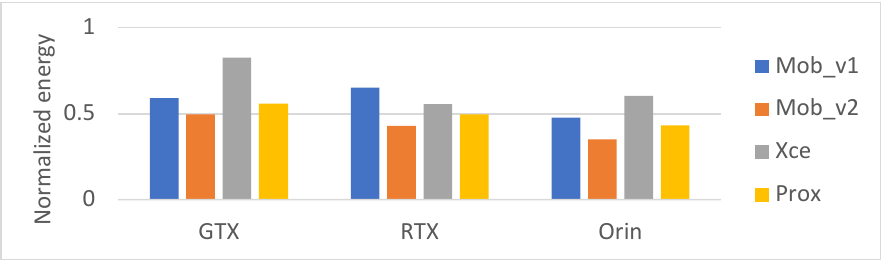}
         \caption{Energy per inference using INT8 normalized to TVM}
         \label{fig:tvm_energy_int8}
     \end{subfigure}
\caption{\centering Energy efficiency of CNN implementations using FCMs and FusePlanner suggested LBL kernels compared to TVM}
    \label{fig:tvm_energy}
    
\end{figure*} 

\subsection{End-to-end comparison with TVM}
\label{sec:vs_tvm}

This section compares end-to-end CNN implementations based on the proposed FCMs and FusePlanner-suggested LBL kernels to TVM. In our end-to-end implementations, FusePlanner iterates over the models' DAGs and suggests which layers to fuse and the tiling parameters that minimize the global memory access for both FCMs and LBL kernels (Section~\ref{sec:fuseestimator}). Then the CNNs are implemented accordingly. The fused layers range from 46-58\% of the convolutional layers of the four CNNs.
 
Figure~\ref{fig:tvm_speedup_fp32} and~\ref{fig:tvm_speedup_int8} show the speedup of our implementations over TVM for the four CNNs using FP32 and INT8. Our implementations constantly outperform TVM, achieving maximum speedups of $1.6x$ and $1.8x$ and average speedups of $1.4x$ and $1.5x$ using FP32 and INT8, respectively. Different DNN GPU combinations have different speedups, but Mob\_v1 has, on average, the highest speedup. Mob\_v1 has a simple linear structure, but TVM graph optimizations are more impactful for DNNs with complex DAGs.

Figure~\ref{fig:tvm_energy_fp32} and~\ref{fig:tvm_energy_int8} show the benefit of our implementations on energy efficiency, \textit{i.e.} energy per inference. They show the energy-per-inference of our implementations normalized to that of TVM. On average, our implementations consume $0.59$ and $0.54$ of the energy consumed by TVM using FP32 and INT8 respectively. Using FP32, The lowest energy consumption is $0.34$ of TVM's, which is the case of Mob\_v1 on Orin. Using INT8, the lowest is $0.35$ of TVM's in the case of Mob\_v2 on Orin. Generally speaking, RTX and Orin have higher energy savings compared to GTX. An important observation is that energy savings are, on average, higher than the reduction in running time. This suggests that even when fusion does not improve the latency considerably, e.g. in cases of compute-bound convolutions, reducing the global memory access is still beneficial as it reduces energy consumption.

%% file: sections/7_related_work.tex
\section{Related work}

\emph{Layer-fusion} is a key inter-layer optimization in many state-of-the-art DNN accelerators~\cite{alwani2016fused,shen2017maximizing,xiao2017exploring,wei2018tgpa,xing2019dnnvm,gao2019tangram,zheng2020efficient,cai2021optimus,qararyah2024efficient,jeong2023pin,yang2023isosceles}. It enables processing the intermediate results immediately, which eliminates the need to frequently access main memory~\cite{alwani2016fused,cai2021optimus,jeong2023pin}. Fusion is also used to maintain high throughput on heterogeneous accelerators that process different CNN layers using multiple layer-custom engines~\cite{umuroglu2017finn, qararyah2024efficient, xiao2017exploring, qararyah2022fibha, shen2017maximizing}. In sparse DNNs, where fewer effectual operations and data-reuse opportunities are present within a layer, fusion maintains a reasonable efficiency by offering higher levels of reuse, and resource utilization~\cite{yang2023isosceles}.

The prior art has demonstrated the advantages of layer fusion on DNN accelerators. However, each has its own shortcomings. Alwani \textit{et al.}~\cite{alwani2016fused} and Xiao \textit{et al.}~\cite{xiao2017exploring} proposals handle only linear CNNs like AlexNet and VGG~\cite{krizhevsky2017imagenet,simonyan2014very}. Zheng \textit{et al.}~\cite{zheng2020efficient}, Zhuang et al.~\cite{zhuang2021convolutional}, and Jeong \textit{et al.}~\cite{jeong2023pin} assume that fusion suffers from an inherent limitation which is the amount of redundant computations that scale with the number of fused layers. However, others~\cite{alwani2016fused, cai2021optimus} have shown that redundant computations can be avoided at the cost of affordable extra buffering. Xing \textit{et al.}~\cite{xing2019dnnvm} and Wei \textit{et al.}~\cite{wei2018tgpa} optimize the execution time but do not consider minimizing DRAM accesses as a main objective. Olyaiy \textit{et al.}~\cite{olyaiy2021accelerating} proposed a fusing technique that targets the bottleneck block structures, the proposed technique reduces the multiplications by up to $20x$ at the cost of extra additions. FINN variants~\cite{umuroglu2017finn, blott2018finn} focus on aggressively quantized models with binary or ternary weights and intermediate results. Yang \textit{et al.}~\cite{yang2023isosceles} fuse layers of highly-sparse models. Other fusion and pipelining proposals, e.g. ~\cite{li2016high,shen2017maximizing, gao2019tangram}, work at the granularity of different inputs or batches. Working at such high granularity increases the inference latency and the off-chip traffic. Convfusion~\cite{waeijen2021convfusion} proposes hardware-agnostic fusion but leaves supporting DW convolution and SIMD to future work.

On GPUs, the most common forms of fusion fall under the categories described by TVM authors~\cite{chen2018tvm}. First, multiple \emph{injective}, or one-to-one e.g. \emph{add} operators, are fused. Second, an \emph{injective} operator is fused with a \emph{reduction} operator. Third, the convolution operator is fused with one or more element-wise operators like normalization and non-linearity. Jia \textit{et al.}~\cite{jia2020enabling} propose a technique to fuse stages of the Winograd convolution algorithm. Li \textit{et al.}~\cite{li2016optimizing} propose to fuse softmax layer implementation to reduce its memory accesses. Chimera~\cite{zheng2023chimera} fuses multiple convolutions on GPUs but does not support DW and the modeling of inter-block optimizations and data movement estimations don’t directly apply to DW.

Unlike the prior art, in this work, we explore fusing DW and PW convolutions to overcome memory access bottlenecks on GPUs. We identify the fusion challenges and trade-offs given the GPU architecture and propose cost models and a set of fused kernels that minimize these convolutions' global memory accesses leading to low latency and energy-efficient inference.


%% file: sections/8_conclusion.tex
\section{Conclusion}
\label{sec:conclusion}

Depthwise and pointwise convolutions are used to design compact DNNs. However, they have a lower compute-to-memory access ratio than the standard convolution, making their global memory access often a bottleneck. This paper proposes fusion as a technique to reduce these convolutions' global memory accesses on GPUs leading to improvements in their efficiency. We propose a set of novel fused convolutional modules (FCMs), GPU kernels composed of fused depthwise and pointwise convolutions. We also propose FusePlanner which consists of cost models to estimate global memory access of layer-by-layer and FCM kernels. Given a GPU architecture, FusePlanner decides when to fuse, and which FCMs to use. Our experiments show that FCMs achieve up to $1.8x$ speedup over a layer-by-layer implementation and up to $3.7x$ over cuDNN. End-to-end implementations of four CNNs using the proposed kernels achieve up to 1.8x speedup compared to TVM-optimized models and consume as little as $34\%$ TVM-optimized models energy.